# Real-Time Magnetometry Using Dark States of a Nitrogen Vacancy Center

Ethan Turner, Shu-Hao Wu, Xinzhu Li, and Hailin Wang

Department of Physics, University of Oregon, Eugene, Oregon 97403, USA

Abstract

We demonstrate real-time magnetometry by detecting fluorescence from a nitrogen vacancy center in the setting of coherent population trapping and by estimating magnetic field from the time series of the observed photon counts, which are correlated with the underlying field. The proof-of-principle experiment uses an external time-varying magnetic field that follows an Ornstein-Uhlenbeck (OU) process. By taking into consideration the statistical properties of the OU process, a Bayesian inference-based estimator can effectively update dynamical information of the field in real time with the detection of just a single photon.

Quantum sensors using a single solid-state spin, such as a negatively charged nitrogen vacancy (NV) center in diamond, can enable the sensing of magnetic fields, electric fields, temperature, and strain with a remarkable combination of high sensitivity and nanometer spatial resolution [1-5]. While experimental studies in quantum sensing have traditionally focused on the measurement of static as well as periodic signals [6,7], detections of time-varying signals have also attracted increasing experimental efforts [8-10]. Real-time sensing with a single spin can open a new frontier for exploring quantum dynamics, quantum fluctuations, and feedback control at the nanoscale.

Thus far, nearly all the sensing experiments with single spins have been based on the use of Ramsey interferometry [1,11]. An individual Ramsey interferometric measurement consists of three sequential stages: initialization, free precession, and read-out. The Ramsey fringes observed probe the free precession of a single or a collection of spins. Repeated Ramsey interferometric measurements provide information on the spin system only in the specific and limited time intervals. In this regard, there are considerable limitations to using Ramsey interferometry for real-time sensing.

Here, we report the experimental demonstration of real-time sensing of a fluctuating magnetic field by exploiting coherent population trapping (CPT) of a NV center. For the CPT process[12-16], the NV center is prepared in a special superposition of two spin states, i.e., the dark state, which prevents optical excitation and emission through destructive quantum interference [17]. A time-varying magnetic field can kick the NV center out of the dark state, leading to a sequence of single-photon emissions from the NV center. We have used this sequence of single-photon emissions to estimate the time-varying magnetic field in real time, with the estimation carried out in a field programmable gate array (FPGA), as illustrated in Fig. 1.

Experimentally, the detection rate of the single-photon emissions limits the effective updating rate of the real-time estimation. Since the overall collection/detection efficiency for optical emissions from a NV center is only a few percent under typical experimental conditions, a key challenge is to obtain dynamical information with the few photons detected. We show that by taking advantage of the statistical properties of the time-varying fields, we can use Bayesian inference to update dynamical information in real time with the detection of just a single photon, demonstrating real-time magnetometry with single detected photons. A detailed analysis further shows good agreement between the experiment and the theoretical expectation. The experimental results are also compared with the classical Cramer-Rao lower bound (CRLB) for the estimation process.

For real-time magnetic-field sensing, we couple the $m_s$=0 and $m_s$=+1 ground spin states to the $E_y$ excited state in a NV center through two dipole optical transitions driven respectively by two resonant optical fields (see Fig. 1). In the limit of equal Rabi frequency, $\Omega$, the dark state for the $\Lambda$-type three-level system can be simply written as $(|m_s=0>-|m_s=+1>)/\sqrt{2}$. As illustrated in Fig. 1, the electron becomes trapped in the dark state and the optical emission is quenched when the Raman resonance condition $\Delta = \delta - \omega_B = 0$ is satisfied, where $\omega_B$ is the frequency separation between the two spin states and $\delta$ is the detuning between the two laser fields.

A time-varying magnetic field leads to a corresponding change in $\omega_B$, which can kick the NV center out of the dark state. The resulting optical excitation of the NV center gives rise to single-photon emissions. Note that it takes only a few spontaneous emission events for the CPT process to reach steady state [18]. For field variations with a timescale that is long compared with the NV radiative lifetime (12 ns [19]), the time series of photon counts, $\{\mathbf{y}_n\} = \{y_1, y_2, ..., y_n, ...\}$, where $y_n$ is the number of photons detected during the $n$th time interval, each with a duration of $\tau$, carries the information on the corresponding change in $\omega_B$, denoted as $\{\mathbf{x}_n\} = \{x_1, x_2, ..., x_n, ...\}$ [20].

We have used three different estimators to output a time series of estimated frequency changes, $\{\tilde{\mathbf{x}}_n\} = \{\tilde{x}_1, \tilde{x}_2, ..., \tilde{x}_n, ...\}$, from the observed time series of photon counts, $\{\mathbf{y}_n\}$. Bayesian inference, which has been used extensively in earlier sensing studies with NV centers[10,21], follows the Bayes update rule given by,

$$p(x_n | y_n, y_{n-1}, ..., y_1) \propto p_{\bar{y}_n}(y_n | x_n) \times p'(x_n | y_{n-1}, ..., y_1), \qquad (1)$$

where $p'(x_n | y_{n-1}, ..., y_1)$ is the prior probability distribution based on the previous time series of photon counts, $p(x_n | y_n, y_{n-1}, ..., y_1)$ is the posteriori probability distribution, and $p_{\bar{y}_n}(y_n | x_n)$ is the likelihood of detecting $y_n$ photons in the $n$th time interval given $x_n$. We assume that $p_{\bar{y}_n}(y_n | x_n)$ follows a Poisson distribution,

$$p_{\bar{y}_n}(y_n | x_n) = \frac{\bar{y}_n^{y_n} e^{-\bar{y}_n}}{y_n!}, \qquad (2)$$

where $\bar{y}_n \propto \tau \rho_{ee}(x_n)$ is the expected average photon count per updating time interval, with $\rho_{ee}$ being the excited-state population. The estimation is then given by

$$\tilde{x}(t) = \int p(x,t) x \, dx. \tag{3}$$

To achieve maximum time-resolution, we usually have $\bar{y}_n < 1$.

The prior probability distribution in Eq. 1 can be significantly improved if the statistical properties of the time-varying field is known. A variety of time-varying processes can be modeled as an Ornstein-Uhlenbeck (OU) process [22-25], which is both Gaussian and Markovian and features an autocorrelation function given by

$$R(t) = \langle x(t_0) x(t_0 + t) \rangle = \sigma^2 e^{-|t|/\tau_c}, \tag{4}$$

with $\sigma^2$ and $\tau_c$ being the variance and correlation time, respectively. Details of the improved prior, which takes into account the statistical properties of the OU process, are discussed in [26] as well as in an earlier theoretical study [20].

In addition to the Bayesian inference, a more intuitive approach is to simply use photon counts detected in a relatively long duration, $\tau_a$, to estimate $x(t)$ by using

$$y_n^{(a)} \propto \tau_a \rho_{ee}(\tilde{x}_n), \tag{5}$$

where $y_n^{(a)}$ is the photon count accumulated between time $n\tau - \tau_a$ and $n\tau$. To achieve an acceptable photon count (of order 10 or greater) for this average count estimator, we need $\tau_a \gg \tau$. For the experimental results presented in this paper, we took $\tau_a = 1.4 \tau_c$ [26].

Our experimental studies were carried out at 10 K with an electronic grade chemical-vapor-deposition grown diamond sample from Element Six. A solid immersion lens (SIL) fabricated on the diamond sample along with a confocal optical microscopy setup was used for the optical excitation and fluorescence collection of a single NV[15,27]. A permanent magnet was used to split the $m_s = \pm 1$ states by 590 MHz. The two resonant optical fields for the CPT process, with an estimated $\Omega/2\pi$ of order 5 MHz, were derived from a 637 nm diode laser and a sideband generated by an electro-optical modulator (EOM). Under these conditions, the CPT dip obtained from the Λ-type system depicted in Fig. 1 features a linewidth of 11.6 MHz, which includes contributions from hyperfine splitting (2.2 MHz), spin dephasing (0.6 MHz) as well as power broadening (near 5 MHz) [26].

For a proof-of-principle demonstration, we apply an external time-varying magnetic field to the NV center by passing an electric current through the coplanar waveguide (CPW) fabricated next to the SIL. The electrical current, which is generated by an arbitrary function generator (AFG), follows a simulated OU process with given $\sigma$ and $\tau_c$ and with $<x(t)>=0$. The use of an external field has enabled us to investigate the dependence of CPT-based real-time sensing on key parameters such as $\sigma$ and $\tau_c$.

For the estimation experiment, the NV center is first initialized to the $m_s$=0 ground state by a 10 μs green laser pulse ($\lambda$=532 nm). This is followed by the application of two resonant optical fields for CPT and the detection of fluorescence from the NV center. To avoid NV ionization due to the resonant optical excitation, we limit the CPT and the fluorescence detection to a duration of 100 μs before reinitializing the NV with a 10 μs green laser pulse. Numerical calculations of the real-time estimation are carried out in a FPGA in a Keysight M3302A card, which also contains a digitizer and an arbitrary waveform generator (AWG). The digitizer tallies the photon counts per update time interval and the AWG outputs the corresponding estimation. The overall process for generating a single update takes about 7 μs. We thus set the updating time interval to be $\tau$=10 μs. Details of the experimental setup are presented in the supplement [26].

Figures 2a and 2b show, as an example, estimations obtained with the OU-Bayesian estimator (which takes into account the statistical properties of the OU process) and with the average count estimator, respectively, as well as a direct comparison between the estimations and the actual frequency changes. For these experiments, we used $\sigma/2\pi$=2.2 MHz, $\tau_c$=5 ms, and an average photon count rate of 5400 per second. The Raman detuning or the bias was set to $(\delta-<\omega_B>)/2\pi=$ 4 MHz (the choice of the bias will be discussed in detail later). As can be seen from these figures, estimations obtained with the OU-Bayesian estimator closely follow the actual field dynamics, whereas estimations obtained with the average count estimator exhibit large deviations from the actual frequency changes for extended periods of time.

To further highlight the differences between these two estimators, we compare estimations obtained in relatively short durations marked by the dashed-line boxes in Figs. 2a and 2b with the corresponding time series of photon counts. As shown in Fig. 2c, the OU-Bayesian estimator can effectively update the dynamics of the frequency change in real time with the detection of just a single photon. Note that an earlier study has used the complete CPT spectrum of a single NV for the sensing of the magnetic fluctuations induced by the nuclear spin bath [13], for which it takes about 100 detected photons to obtain a single estimation.

For the average count estimator, a relatively large number of photons need to be detected in order to appreciably change the estimations, as can be seen from Fig. 2d. There is also a large delay between the estimation and the actual frequency change due to the relatively long $\tau_a$ used. Note that significantly reducing $\tau_a$ and thus $y_n^{(a)}$ only leads to greater fluctuations in $(\tilde{x}_n - x_n)$. Average count estimators, while performing poorly for real-time sensing, work well for the sensing of static signals. For example, electromagnetically induced transparency (EIT), which is closely related to CPT, of an ensemble of NV centers has been used successfully for static sensing [28].

For a quantitative analysis of the estimations, we have examined the estimation variance defined as $\mathrm{Var}[\tilde{x}_n] = \langle (\tilde{x}_n - x_n)^2 \rangle$ and in particular $\langle (\tilde{x}_n - x_n)^2 \rangle / \sigma^2$, denoted as $\mathrm{Var}/\sigma^2$. We have also carried out detailed comparisons between the estimation variances obtained from the experimentally observed time series of photon counts and those obtained from the theoretically simulated time series of photon counts. The theoretical model used for the simulations has been presented in our earlier study [20] and is also discussed in detail in the supplement [26]. For sensing of a time-varying signal with a given distribution, the estimation performance averaged over the entire distribution rather than that at a single point is important. In this regard, the sensing sensitivity as defined for static sensing is no longer applicable. Instead, we use the estimation variance as an effective measure of the sensing performance [29].

Figure 3a plots $\mathrm{Var}/\sigma^2$ as a function of $\tau_c$ and compares the relative variances obtained with the OU-Bayesian estimator, the average count estimator, and the simple Bayesian estimator, which takes no account of the statistical properties of the OU process. As expected, the variances for both the average count and the OU-Bayesian estimators decrease with increasing $\tau_c$. Nevertheless, $\mathrm{Var}/\sigma^2$ for the average count estimator is far above 1 when $\tau_c$ is near 1 ms and only falls slightly below 1 when $\tau_c$ approaches 10 ms. In comparison, $\mathrm{Var}/\sigma^2$ for the OU-Bayesian estimator remains significantly below 1 when $\tau_c$ approaches 1 ms.

It should be noted that with $\mathrm{Var}/\sigma^2 \approx 1$ for the range of $\tau_c$ used in Fig. 3a, the simple Bayesian estimator essentially provides no information on the time-varying field. As shown in our earlier theoretical study [20] and confirmed in additional experiments, estimations obtained with the simple Bayesian estimator quickly converge to the average value, with $\tilde{x}(t) \approx 0$.

Figure 3b compares the experimentally observed variances with the corresponding simulated variances. Both variances were obtained with the OU-Bayesian estimator. We found that the experimentally observed variances are considerably greater than the simulated variances, for which a perfect charge initialization for the negatively charged NV center is assumed. A detailed analysis shows that for our experiments, the charge initialization fidelity is about 75% [26]. Including the non-ideal charge initialization in the model yields a good agreement between the experiment and the simulation. Figure 3b also shows that the experimentally observed variance is considerably above the calculated CRLB (see [20] for a detailed discussion on the calculation of the CRLB). Theoretically, CRLB can be reached only when the relevant CPT spectral response is linear or quadratic, which is not the case for the actual experiment.

The estimation variances depend on the choice of CPT parameters, especially the bias. Figure 4a shows $\text{Var}/\sigma^2$ obtained with the OU-Bayesian estimator as a function of the bias, with other experimental conditions the same as those used for Fig. 2a. As expected, the estimations become ineffective when the bias approaches 0 (i.e., near the bottom of the CPT dip), in agreement with the theoretical expectation. The estimations also perform poorly when the bias significantly exceeds the half width of the CPT dip. Note that near the wings of the CPT dip, effects of hyperfine splitting of the relevant spin states, which are not included in the theoretical model, become important, leading to the observed variances that are larger than the simulated values, as shown in Fig. 4a.

The sensitivity and range of the real-time sensing process also depend on the CPT parameters. In particular, there is a trade-off between the largest and the smallest frequency changes that can be sensed via a CPT process. The smaller the CPT linewidth, the more sensitive the CPT-based sensing process becomes, whereas the CPT linewidth sets the range of the sensing process. Figure 4b shows $\text{Var}/\sigma^2$ obtained with the OU-Bayesian estimator as a function of $\sigma$, with other experimental conditions the same as those used for Fig. 2a. For relatively small $\sigma$, $\text{Var}/\sigma^2$ increases with decreasing $\sigma$ and goes above 1 when $\sigma/2\pi$ falls below 0.5 MHz. In this case, the large CPT linewidth (11.6 MHz as mentioned earlier) used in the experiment limits the sensitivity of the real-time sensing process. The experimental results are in good agreement with the simulated values, as shown in Fig. 4b.

We can further improve the sensitivity of the real-time sensing process by reducing the CPT linewidth. For example, polarizing the $^{14}$N nuclear spin with optical pumping avoids the complication of hyperfine splitting [30]. The use of isotopically purified diamond reduces dephasing induced by the nuclear spin bath and can lead to a CPT linewidth less than 1 kHz [31]. Additional improvements in the overall sensing performance can also be achieved through better charge initialization, for example, through the use of real-time control techniques [32].

In summary, we have demonstrated real-time magnetometry using dark states in a NV center by estimating the time-varying magnetic field from the corresponding time series of photon counts in a CPT setting. A Bayesian estimator, which takes advantage of the statistical properties of the time-varying field, can effectively update the dynamical information of the field with the detection of a single photon, which is otherwise not feasible with the more conventional average-count estimator that takes no account of the statistical properties. While a NV center has been used as a model system for the proof-of-principle demonstration, the real-time magnetometry can also be extended to other solid-state spin systems [33].

Real-time magnetometry using a single solid-state spin can add a new and powerful tool to quantum sensing. The magnetometry can be used for studies of time-varying magnetic fields in a variety of systems at the nanoscale, for example, nuclear spin baths [13,20,34] and two-dimensional semiconductors [35,36]. Combining real-time sensing with feedback control also opens new avenues, such as protecting a spin qubit from the fluctuating magnetic environment via feedback control [9,13].

This work is supported by the ARO MURI Grant No. W911NF-18-1-0218.

E.T. and S.W. contributed equally to this work.

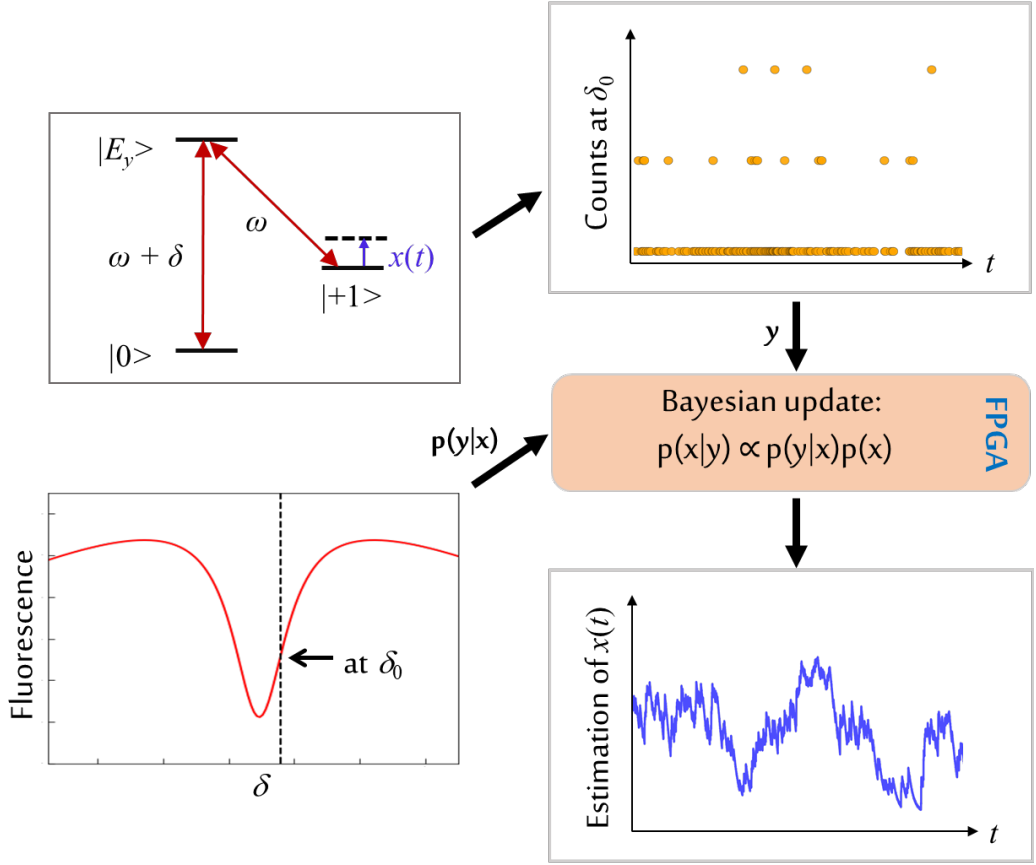

FIG. 1. (color online) Schematic illustrating real-time sensing of magnetic fields using single-photon emissions from a NV center prepared in a CPT setting. Bayesian inference is used to estimate time-varying magnetic fields from the corresponding time series of photon counts. The upper left figure shows the Λ-type three-level system used for the CPT process. The lower left figure shows schematically the fluorescence from the excited state as a function of the detuning between the two applied optical fields. The fluorescence is quenched when the detuning equals the frequency separation between the two lower states.

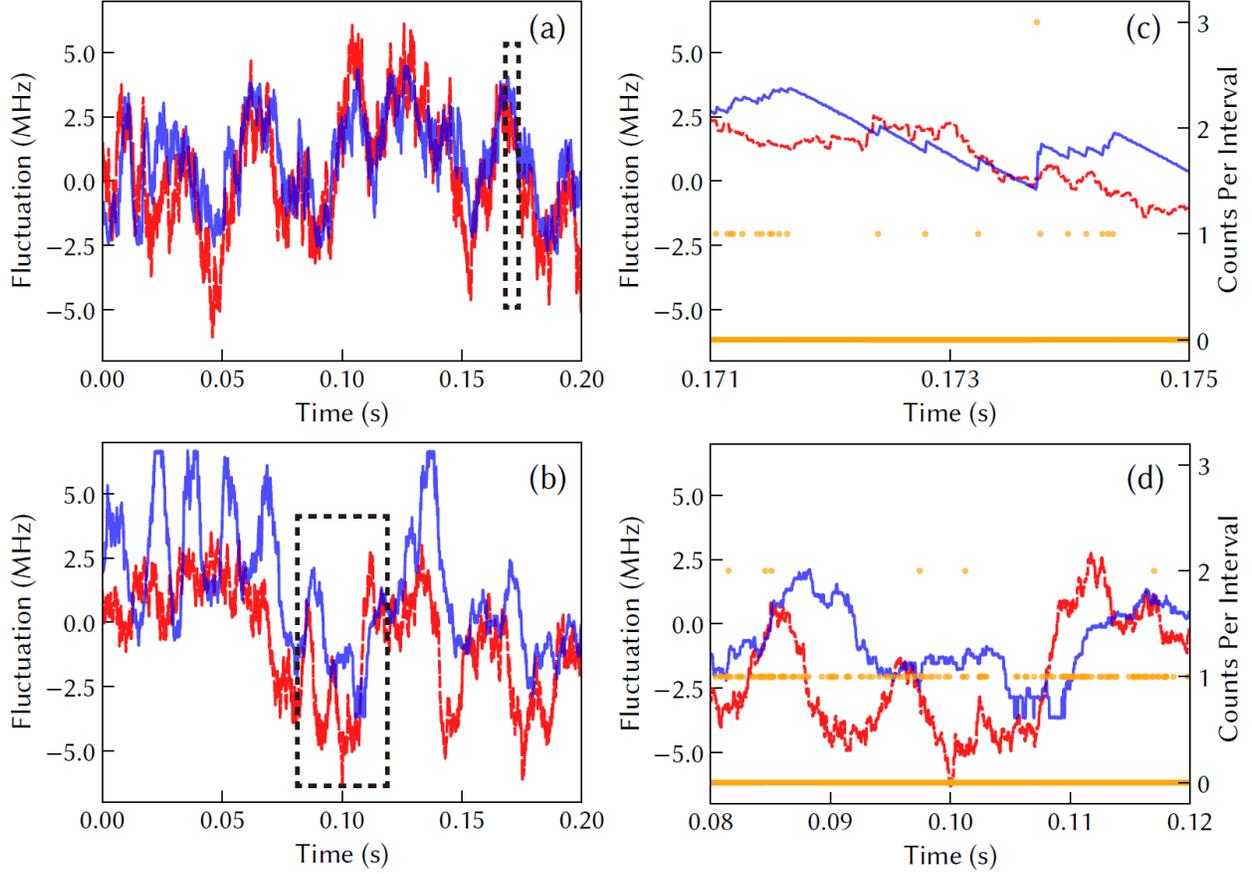

FIG. 2. (color online) (a) and (b) Estimations (blue solid line) of the fluctuations in $\omega_B/2\pi$ obtained with the OU-Bayesian and the average count estimators, respectively, along with the actual fluctuations (red dashed line). (c) and (d) A closer look at the results in the dashed-line boxes in (a) and (b), respectively, along with the corresponding photon counts (orange dots) per updating interval.

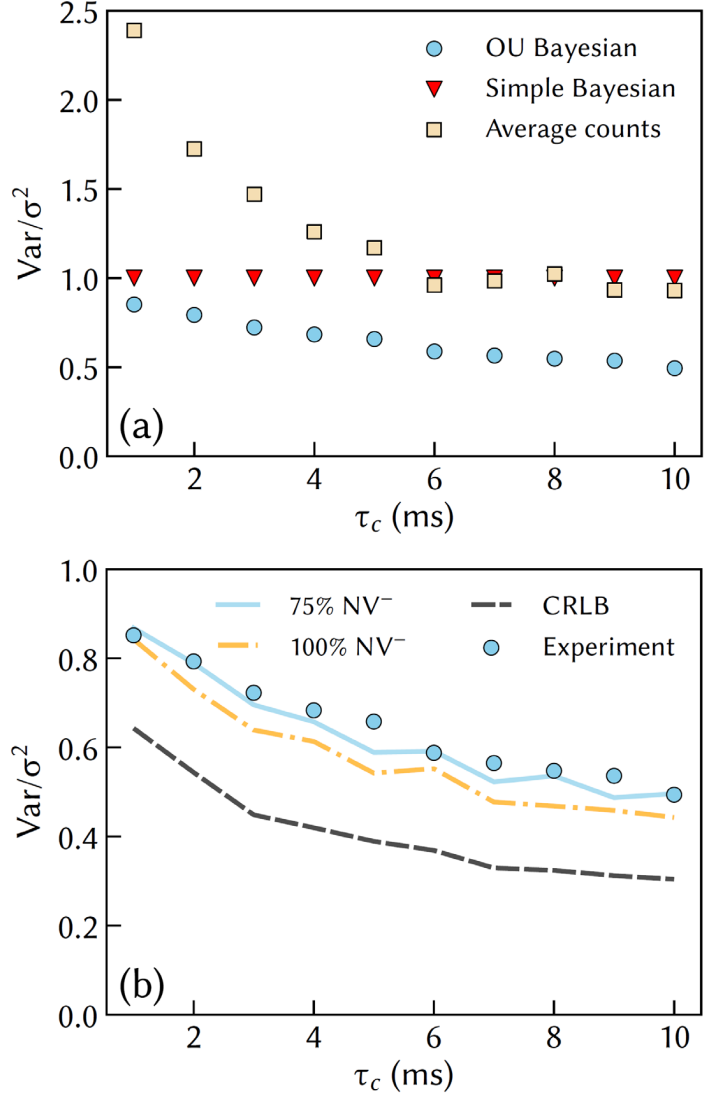

FIG. 3. (colon online) (a) Comparison of estimation variances obtained with OU-Bayesian, simple Bayesian, and average count estimators as a function of $\tau_c$. (b) Comparison of the estimation variances obtained with the OU-Bayesian estimator with the corresponding simulated values, for which a charge initialization fidelity of 100% (dotted line) and 75% (solid line) is assumed. The dashed line shows the calculated CRLB. Experimental parameters used are the same as those for Fig. 2 unless otherwise specified.

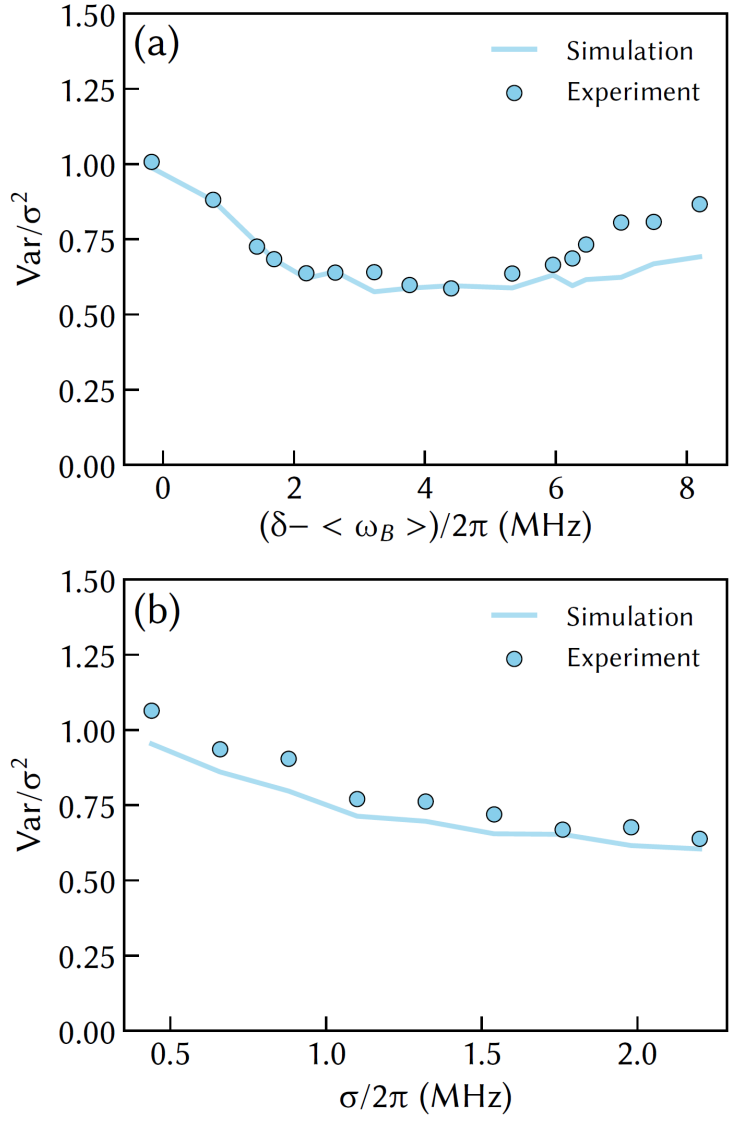

FIG. 4. (color online) (a) Estimation variances obtained with the OU-Bayesian estimator as a function of the bias (i.e., Raman detuning). (b) Estimation variances obtained with the OU-Bayesian estimator as function of $\sigma$. The solid lines in both figures show the corresponding simulated values.

**References:**

[1] J. M. Taylor, P. Cappellaro, L. Childress, L. Jiang, D. Budker, P. R. Hemmer, A. Yacoby, R. Walsworth, and M. D. Lukin, *High-sensitivity diamond magnetometer with nanoscale resolution*, Nature Physics **4**, 810 (2008).

[2] G. Balasubramanian, I. Y. Chan, R. Kolesov, M. Al-Hmoud, J. Tisler, C. Shin, C. Kim, A. Wojcik, P. R. Hemmer, A. Krueger, T. Hanke, A. Leitenstorfer, R. Bratschitsch, F. Jelezko, and J. Wrachtrup, *Nanoscale imaging magnetometry with diamond spins under ambient conditions*, Nature **455**, 648 (2008).

[3] F. Dolde, H. Fedder, M. W. Doherty, T. Nobauer, F. Rempp, G. Balasubramanian, T. Wolf, F. Reinhard, L. C. L. Hollenberg, F. Jelezko, and J. Wrachtrup, *Electric-field sensing using single diamond spins*, Nature Physics **7**, 459 (2011).

[4] G. Kucsko, P. C. Maurer, N. Y. Yao, M. Kubo, H. J. Noh, P. K. Lo, H. Park, and M. D. Lukin, *Nanometre-scale thermometry in a living cell*, Nature **500**, 54 (2013).

[5] M. W. Doherty, V. V. Struzhkin, D. A. Simpson, L. P. McGuinness, Y. F. Meng, A. Stacey, T. J. Karle, R. J. Hemley, N. B. Manson, L. C. L. Hollenberg, and S. Prawer, *Electronic Properties and Metrology Applications of the Diamond NV- Center under Pressure*, Physical Review Letters **112**, 047601 (2014).

[6] C. L. Degen, F. Reinhard, and P. Cappellaro, *Quantum sensing*, Rev Mod Phys **89**, 035002 (2017).

[7] V. Giovannetti, S. Lloyd, and L. Maccone, *Advances in quantum metrology*, Nat Photonics **5**, 222 (2011).

[8] A. Cooper, E. Magesan, H. N. Yum, and P. Cappellaro, *Time-resolved magnetic sensing with electronic spins in diamond*, Nat Commun **5**, 3141 (2014).

[9] M. D. Shulman, S. P. Harvey, J. M. Nichol, S. D. Bartlett, A. C. Doherty, V. Umansky, and A. Yacoby, *Suppressing qubit dephasing using real-time Hamiltonian estimation*, Nat Commun **5**, 5156 (2014).

[10] R. Santagati, A. A. Gentile, S. Knauer, S. Schmitt, S. Paesani, C. Granade, N. Wiebe, C. Osterkamp, L. P. McGuinness, J. Wang, M. G. Thompson, J. G. Rarity, F. Jelezko, and A. Laing, *Magnetic-Field Learning Using a Single Electronic Spin in Diamond with One-Photon Readout at Room Temperature*, Phys Rev X **9**, 021019 (2019).

[11] H. Lee, P. Kok, and J. P. Dowling, *A quantum Rosetta stone for interferometry*, J Mod Optic **49**, 2325 (2002).

[12] C. Santori, P. Tamarat, P. Neumann, J. Wrachtrup, D. Fattal, R. G. Beausoleil, J. Rabeau, P. Olivero, A. D. Greentree, S. Prawer, F. Jelezko, and P. Hemmer, *Coherent population trapping of single spins in diamond under optical excitation*, Physical Review Letters **97**, 247401 (2006).

[13] E. Togan, Y. Chu, A. Imamoglu, and M. D. Lukin, *Laser cooling and real-time measurement of the nuclear spin environment of a solid-state qubit*, Nature **478**, 497 (2011).

[14] C. G. Yale, B. B. Buckley, D. J. Christle, G. Burkard, F. J. Heremans, L. C. Bassett, and D. D. Awschalom, *All-optical control of a solid-state spin using coherent dark states*, Proceedings of the National Academy of Sciences of the United States of America **110**, 7595 (2013).

[15] D. A. Golter, T. K. Baldwin, and H. L. Wang, *Protecting a Solid-State Spin from Decoherence Using Dressed Spin States*, Physical Review Letters **113**, 237601 (2014).

[16] D. A. Golter, T. Oo, M. Amezcua, I. Lekavicius, K. A. Stewart, and H. Wang, *Coupling a Surface Acoustic Wave to an Electron Spin in Diamond via a Dark State*, Physical Review X **6**, 041060 (2016).

[17] M. O. Scully and M. S. Zubairy, *Quantum optics* (Cambridge University Press, 1997).

[18]  I. Lekavicius, D. A. Golter, T. Oo, and H. L. Wang, *Transfer of Phase Information between Microwave and Optical Fields via an Electron Spin*, Physical Review Letters **119**, 063601 (2017).
[19]  P. Tamarat, T. Gaebel, J. R. Rabeau, M. Khan, A. D. Greentree, H. Wilson, L. C. L. Hollenberg, S. Prawer, P. Hemmer, F. Jelezko, and J. Wrachtrup, *Stark shift control of single optical centers in diamond*, Physical Review Letters **97**, 083002 (2006).
[20]  S. H. Wu, E. Turner, and H. L. Wang, *Continuous real-time sensing with a nitrogen-vacancy center via coherent population trapping*, Phys Rev A **103**, 042607 (2021).
[21]  C. Bonato, M. S. Blok, H. T. Dinani, D. W. Berry, M. L. Markham, D. J. Twitchen, and R. Hanson, *Optimized quantum sensing with a single electron spin using real-time adaptive measurements*, Nat Nanotechnol **11**, 247 (2016).
[22]  V. V. Dobrovitski, A. E. Feiguin, R. Hanson, and D. D. Awschalom, *Decay of Rabi Oscillations by Dipolar-Coupled Dynamical Spin Environments*, Phys Rev Lett **102** (2009).
[23]  W. M. Witzel, M. S. Carroll, L. Cywinski, and S. Das Sarma, *Quantum decoherence of the central spin in a sparse system of dipolar coupled spins*, Phys Rev B **86**, 035452 (2012).
[24]  G. de Lange, Z. H. Wang, D. Riste, V. V. Dobrovitski, and R. Hanson, *Universal Dynamical Decoupling of a Single Solid-State Spin from a Spin Bath*, Science **330**, 60 (2010).
[25]  C. Zhang and K. Molmer, *Estimating a fluctuating magnetic field with a continuously monitored atomic ensemble*, Phys Rev A **102**, 063716 (2020).
[26]  *See the Supplementary Material.*
[27]  D. A. Golter and H. L. Wang, *Optically Driven Rabi Oscillations and Adiabatic Passage of Single Electron Spins in Diamond*, Physical Review Letters **112**, 116403 (2014).
[28]  V. M. Acosta, K. Jensen, C. Santori, D. Budker, and R. G. Beausoleil, *Electromagnetically Induced Transparency in a Diamond Spin Ensemble Enables All-Optical Electromagnetic Field Sensing*, Phys Rev Lett **110**, 213605 (2013).
[29]  C. Bonato and D. W. Berry, *Adaptive tracking of a time-varying field with a quantum sensor*, Phys Rev A **95**, 052348 (2017).
[30]  M. L. Goldman, T. L. Patti, D. Levonian, S. F. Yelin, and M. D. Lukin, *Optical Control of a Single Nuclear Spin in the Solid State*, Phys Rev Lett **124**, 153203 (2020).
[31]  G. Balasubramanian, P. Neumann, D. Twitchen, M. Markham, R. Kolesov, N. Mizuochi, J. Isoya, J. Achard, J. Beck, J. Tissler, V. Jacques, P. R. Hemmer, F. Jelezko, and J. Wrachtrup, *Ultralong spin coherence time in isotopically engineered diamond*, Nature Materials **8**, 383 (2009).
[32]  D. A. Hopper, J. D. Lauigan, T. Y. Huang, and L. C. Bassett, *Real-Time Charge Initialization of Diamond Nitrogen-Vacancy Centers for Enhanced Spin Readout*, Phys Rev Appl **13**, 024016 (2020).
[33]  D. D. Awschalom, R. Hanson, J. Wrachtrup, and B. B. Zhou, *Quantum technologies with optically interfaced solid-state spins*, Nature Photonics **12**, 516 (2018).
[34]  N. Zhao, J. L. Hu, S. W. Ho, J. T. K. Wan, and R. B. Liu, *Atomic-scale magnetometry of distant nuclear spin clusters via nitrogen-vacancy spin in diamond*, Nat Nanotechnol **6**, 242 (2011).
[35]  I. Lovchinsky, J. D. Sanchez-Yamagishi, E. K. Urbach, S. Choi, S. Fang, T. I. Andersen, K. Watanabe, T. Taniguchi, A. Bylinskii, E. Kaxiras, P. Kim, H. Park, and M. D. Lukin, *Magnetic resonance spectroscopy of an atomically thin material using a single-spin qubit*, Science **355**, 503 (2017).
[36]  L. Thiel, Z. Wang, M. A. Tschudin, D. Rohner, I. Gutierrez-Lezama, N. Ubrig, M. Gibertini, E. Giannini, A. F. Morpurgo, and P. Maletinsky, *Probing magnetism in 2D materials at the nanoscale with single-spin microscopy*, Science **364**, 973 (2019).